\documentclass[acmsmall]{acmart}
\usepackage{multirow}
\usepackage{booktabs}
\usepackage{enumitem} 

\AtBeginDocument{%
  }

\copyrightyear{2026}
\acmYear{2026}
\setcopyright{cc}
\setcctype{by}
\acmConference[CSCW Companion '26]{Companion of the Computer-Supported Cooperative Work and Social Computing}{October 10--14, 2026}{Salt Lake City, UT, USA}
\acmBooktitle{Companion of the Computer-Supported Cooperative Work and Social Computing (CSCW Companion '26), October 10--14, 2026, Salt Lake City, UT, USA}
\acmDOI{10.1145/3785651.3831510}
\acmISBN{979-8-4007-2378-0/2026/10}

\begin{document}

\title{Octo’s Adventure: At-home Deployment of a Pediatric Education Tool} \enlargethispage*{16pt}

\author{Crimson Olaleye}
\affiliation{%
  \institution{University of Minnesota}
 \city{Minneapolis}
\state{Minnesota}
  \country{USA}}
\email{olale016@umn.edu}

\author{Neda Barbazi}
\affiliation{%
  \institution{University of Minnesota}
 \city{Minneapolis}
  \country{USA}}
\email{barba087@umn.edu}

\author{Ji Youn Shin}
\affiliation{%
  \institution{University of Minnesota}
 \city{Minneapolis}
  \country{USA}}
\email{shinjy@umn.edu}

\author{Gurumurthy Hiremath}
\affiliation{%
  \institution{University of Minnesota}
\city{Minneapolis}
  \country{USA}}
\email{hiremath@umn.edu}

\author{Carlye Anne Lauff}
\affiliation{%
  \institution{University of Minnesota}
\city{Minneapolis}
  \country{USA}}
\email{carlye@umn.edu}

\renewcommand{\shortauthors}{Olaleye et al.}

\begin{abstract}
Children with congenital heart disease (CHD) and their families often navigate educational, emotional, and communication challenges regarding care management outside clinical settings. However, many pediatric health intervention studies rely on caregiver-mediated feedback or structured evaluations, limiting visibility into children’s direct experiences during everyday use. We present a one-week at-home deployment study of Octo, a hybrid physical-digital educational tool designed to support children’s health literacy and reduce parental educational burden. Extending participatory design approaches into deployment, the study embedded child-friendly reflection activities, drawing-based feedback, and emotional tracking into family routines. We conducted the deployment with 13 families of children diagnosed with CHD. Preliminary findings suggest that children gained understanding across Octo’s physical and digital components, expressed emotional attachment and confidence through play, and engaged in collaborative family communication about their condition. These findings will guide future refinement of Octo and inform future clinical deployment across care settings. 
\end{abstract}




\maketitle

\section{Introduction} \enlargethispage*{16pt}
Children with chronic health conditions such as congenital heart disease (CHD), affecting about 1 in 100 newborns globally, often navigate long-term care management that creates emotional, educational, and coordination burdens for children and their families  \cite{CDC2026,Bowers2024,Zhang2021,Barbazi2025Scoping,Barbazi2026DRS}. While existing resources are directed at parents/caregivers, fewer are designed to help children actively understand and engage with their care \cite{Alkan2017,Barbazi2025Scoping,Burns2022,Ma2025,Rodts2020,Zeng2025}. Ethical safeguards, developmental differences, communication styles, and difficulty understanding abstract medical concepts can make children’s direct involvement in their care challenging \cite{Barbazi2025Scoping, Burns2022,Ma2025,Rodts2020,Zeng2025}. To address this gap, we iteratively developed \textit{Octo}, a hybrid physical-digital educational tool, through a multi-phase research study, to support learning through play \cite{Barbazi2025Scoping,Barbazi2025Play,Barbazi2025Assessment,Barbazi2025Boundary, Barbazi2026,Zeng2025}. This pediatric health education tool includes a plush dinosaur with an openable chest, a puzzle heart, a medical-themed bag with play tools (blood pressure cuff, EKG stickers, echo wand, toothbrush), a storybook, and a serious game (SG) digital application \cite{Barbazi2025Boundary,Zeng2025}. Octo aims to support children’s health literacy, emotional expression, and family communication. However, prior phases primarily designed the tool in facilitated settings, leaving its use in everyday family contexts less understood \cite{Barbazi2026,Zeng2025}.

In pediatric care, much of the ongoing learning, emotional processing, and condition-related communication occurs at home between clinical visits \cite{Burns2022,Teela2023,Rodts2020,Barbazi2025Boundary}. Understanding how families use such tools in daily life can reveal whether they fit into routines, sustain engagement, and support learning beyond supervised clinical encounters \cite{Cagiltay2023,Zhang2023,Chaudhury2025}. While prior work has used at-home deployments to examine familial engagement with pediatric technologies in everyday contexts \cite{Park2022,Ullman2021}, many rely on caregiver-mediated feedback, interviews, or post-deployment reports, limiting visibility into children’s direct experiences. This limitation is especially relevant in pediatric care, where parents often act as proxies for children, and their perspectives may be inconsistently captured over time \cite{Cagiltay2023,Moon2025,Ullman2021}. We address this gap by extending participatory design (PD) methods used in early design stages, such as drawing and storytelling \cite{Guha2004,Isola2012}, to the deployment phase through a one-week at-home deployment with 13 families to capture in situ use \cite{Siek2014,Cagiltay2023,Noh2025,Dahlback1993}. This study explores how families interact with Octo and the role it plays in supporting children’s health literacy, emotional empowerment, and parental educational responsibilities. These findings will guide future refinement of Octo’s physical and digital components by identifying usability challenges and incorporating family feedback. Two research questions (RQ) guided our study: \enlargethispage*{16pt}  
\begin{itemize}
    \item RQ1: How does at-home use of a pediatric health education tool shape children’s understanding of and expression about their health conditions?
    \item RQ2: How do parents perceive the role of a pediatric health education tool in supporting children’s understanding of their health conditions and alleviating caregiver burden?
\end{itemize}

\section{Related Work}

\subsection{Healthcare Interventions Designed for Children and Their Families}
HCI and design research increasingly explore technology-mediated healthcare interventions to support children’s understanding of health conditions and help families manage chronic illness in everyday life \cite{Cha2025, barbazi2023colors}. Prior research includes symptom-tracking systems that support shared care routines between children and caregivers, for conditions such as Type 1 diabetes \cite{Cha2024}, family-centered communication tools that help coordinate care across children, caregivers, schools, and healthcare providers \cite{Sepehri2023}, child patient-provider communication technologies \cite{Seo2021}, alongside mobile health applications \cite{Shin2019}, educational games \cite{Nikkila2012,Zeng2025}, conversational agents \cite{Park2022}, and interactive healthcare technologies that help children understand medical conditions, procedures, and healthcare experiences \cite{Isbister2022,Jeong2018}. Other studies explore tangible and play-based healthcare tools that support children’s participation in care while helping families navigate communication, uncertainty, and ongoing healthcare routines in everyday settings \cite{Ahmadpour2023,Barbazi2025Play}. Many of these interventions extend beyond clinical settings into home environments, where children and caregivers integrate them into everyday routines and family practices. Prior work shows that home settings reveal how caregiving responsibilities, family dynamics, and children’s day-to-day behaviors shape technology use over time \cite{Isola2012,Nikkhah2022Care,Thiessen2024}. However, many pediatric interventions rely on caregiver-reported experiences or structured evaluations that may not fully capture children’s direct engagement with interventions in everyday life. Existing research further highlights that children and caregivers often engage with these interventions differently and hold distinct needs and priorities during long-term care routines \cite{Cha2024,Nikkhah2022Adaptive}. Additionally, many interventions receive limited iterative testing in home settings, where family routines and dynamics evolve over time. These challenges highlight the need for approaches that better support meaningful child participation during real-world deployment. \enlargethispage*{16pt}

\subsection{Participatory Approaches with Children During the Design Process}
Participatory design (PD), or co-design with children, is a collaborative approach in which children act as active contributors to the design rather than recipients \cite{Barendregt2016,Druin1999,Kensing1998,Scaife1997}. Since children express themselves in various ways, such as play, drawing, or storytelling \cite{Kender2020}, different strategies have been used to gather participatory feedback from children during the early stages of the design process \cite{Goodacre2025,Guha2004}. During the ideation phase, drawing, collage‑making, and diary probes have been used to allow children to share feelings and ideas visually \cite{Chen2026,Hong2020,OSullivan2021,Silva2024,Farhat2024}, while storytelling has been used during refinement to provide opportunities for imaginative expression \cite{Barbazi2026,Bekker2002,Zhang2022,Carter2005}. Short interviews or play‑based conversations can also provide meaningful insight during refinement, though children may struggle to articulate complex feedback \cite{Jani2025}. In later stages of product development, such as field deployment, the range of PD methods used across HCI narrows. While diary probes and multimodal feedback are common in general field studies \cite{Carter2005}, deployments with children rely on more structured approaches, like interviews, surveys, or emotional logs \cite{Chen2026,Read2006,Stefanidi2025}. These approaches capture valuable insights from children, but lack the expressive and creative participation supported earlier in the design process \cite{Kender2020}. As a result, children’s role as co-designers becomes less visible during real-world deployment \cite{Chisik2020,Druin1999}. We address this gap by extending early-stage PD strategies into Octo's deployment through child-friendly reflection prompts, drawing-based activities, and emotional tracking, designed to minimize parental facilitation \cite{Jani2025,OSullivan2021}. These approaches allow families to provide feedback that fits everyday routines, supports children's participation, and reduces caregiver burden while informing Octo's educational and emotional impact \cite{Stefanidi2025,Thiessen2024}. 

\section{Methods}
We conducted a one-week at-home deployment study on a rolling basis between February and April 2026 with 13 families of children ages 4–10 diagnosed with CHD, with IRB approval from the University of Minnesota (STUDY00020670). Prior research shows that longer pediatric deployments can create participation fatigue \cite{Cha2025}, so we chose a one-week duration based on similar HCI deployment studies \cite{Chaudhury2025,Fu2023,Kim2026,Park2022,Cha2025}. We recruited families through community and clinical settings to capture engagement across these contexts. Eligibility criteria included children ages 4-10, a CHD diagnosis, and English-speaking families (Table~\ref{tab:participant-demographics}). Clinic-based recruitment occurred at the M Health Fairview Explorer Clinic, where a provider introduced the study during routine visits; community recruitment took place at the Feel the Beat Camp Odayin Resource Fair, where Octo prototypes were on display. All interested families completed pre-assessments \cite{Barbazi2025Assessment} and received the deployment kit. \enlargethispage*{16pt}

\begin{table*}[ht] 
\caption{Demographic information of child (C) and parent (P) participants; F indicates the family unit.} 
\label{tab:participant-demographics}

\centering
\small

\resizebox{\textwidth}{!}{%
\begin{tabular}{llllllllllll}
\toprule

\begin{tabular}[t]{@{}l@{}}\textbf{ID\#}\end{tabular} &
\begin{tabular}[t]{@{}l@{}}\textbf{Years}\\\textbf{Diagnosed}\end{tabular} &
\begin{tabular}[t]{@{}l@{}}\textbf{Child}\\\textbf{Age}\end{tabular} &
\begin{tabular}[t]{@{}l@{}}\textbf{Gender}\end{tabular} &
\begin{tabular}[t]{@{}l@{}}\textbf{Race}\end{tabular} &
\begin{tabular}[t]{@{}l@{}}\textbf{Main}\\\textbf{Parent}\end{tabular} &
\begin{tabular}[t]{@{}l@{}}\textbf{Parent}\\\textbf{Age}\end{tabular} &
\begin{tabular}[t]{@{}l@{}}\textbf{Income}\end{tabular} &
\begin{tabular}[t]{@{}l@{}}\textbf{Education}\end{tabular} &
\begin{tabular}[t]{@{}l@{}}\textbf{Race}\end{tabular} &
\begin{tabular}[t]{@{}l@{}}\textbf{Employment}\end{tabular} &
\begin{tabular}[t]{@{}l@{}}\textbf{Marital}\\\textbf{Status}\end{tabular} \\

\midrule

F1 (C1+P1) & Since birth & 9 & M & White & Mom & 35--44 & 100,000+ & Bachelors & White & Employed, full-time & Married \\
F2 (C2+P2) & $<$1 & 9 & M & White & Mom & 35--44 & 25,000--50,000 & Some college & White & Employed, full-time & Single \\
F3 (C3+P3) & Since birth & 9 & F & White & Mom & 35--44 & 100,000+ & Bachelors & White & Employed, full-time & Married \\
F4 (C4+P4) & Since birth & 9 & M & White & Mom & 35--44 & 100,000+ & Bachelor's & White & Employed, full-time & Married \\
F5 (C5+P5) & 1--2 & 5 & M & White & Mom & 25--34 & 100,000+ & Bachelor's & White & Employed, full-time & Married \\
F6 (C6+P6) & Since birth & 8 & F & White & Mom & 45--54 & 100,000+ & Some college & White & Self-employed & Married \\
F7 (C7+P7) & Since birth & 4 & M & White & Mom & 35--44 & Prefer not to say & Doctoral degree & White & Employed, temporary & Married \\
F8 (C8+P8) & Since birth & 5 & M & White & Mom & 35--44 & Prefer not to say & Master's & White & Unemployed & Married \\
F9 (C9+P9) & 3--5 & 5 & F & White & Mom & 35--44 & 100,000+ & Master's & White & Employed, full-time & Married \\
F10 (C10+P10) & Since birth & 10 & M & White & Mom & 35--44 & 100,000+ & Doctoral degree & White & Retired & Married \\
F11 (C11+P11) & 1--2 & 5 & F & White & Mom & 25--34 & 50,000--75,000 & Some college & White & Unemployed & Married \\
F12 (C12+P12) & Since birth & 10 & M & White & Mom & 35--44 & 100,000+ & Bachelors & White & Employed, full-time & Married \\
F13 (C13+P13) & 3--5 & 8 & F & White & Dad & 35--44 & 100,000+ & Masters & White & Employed, full-time & Married \\

\bottomrule
\end{tabular}%
}
\end{table*} \enlargethispage*{10pt}

\subsection{Study Instrument (Deployment Kit) and Procedure}
\begin{figure}[h]
  \centering
  \includegraphics[width=\linewidth]{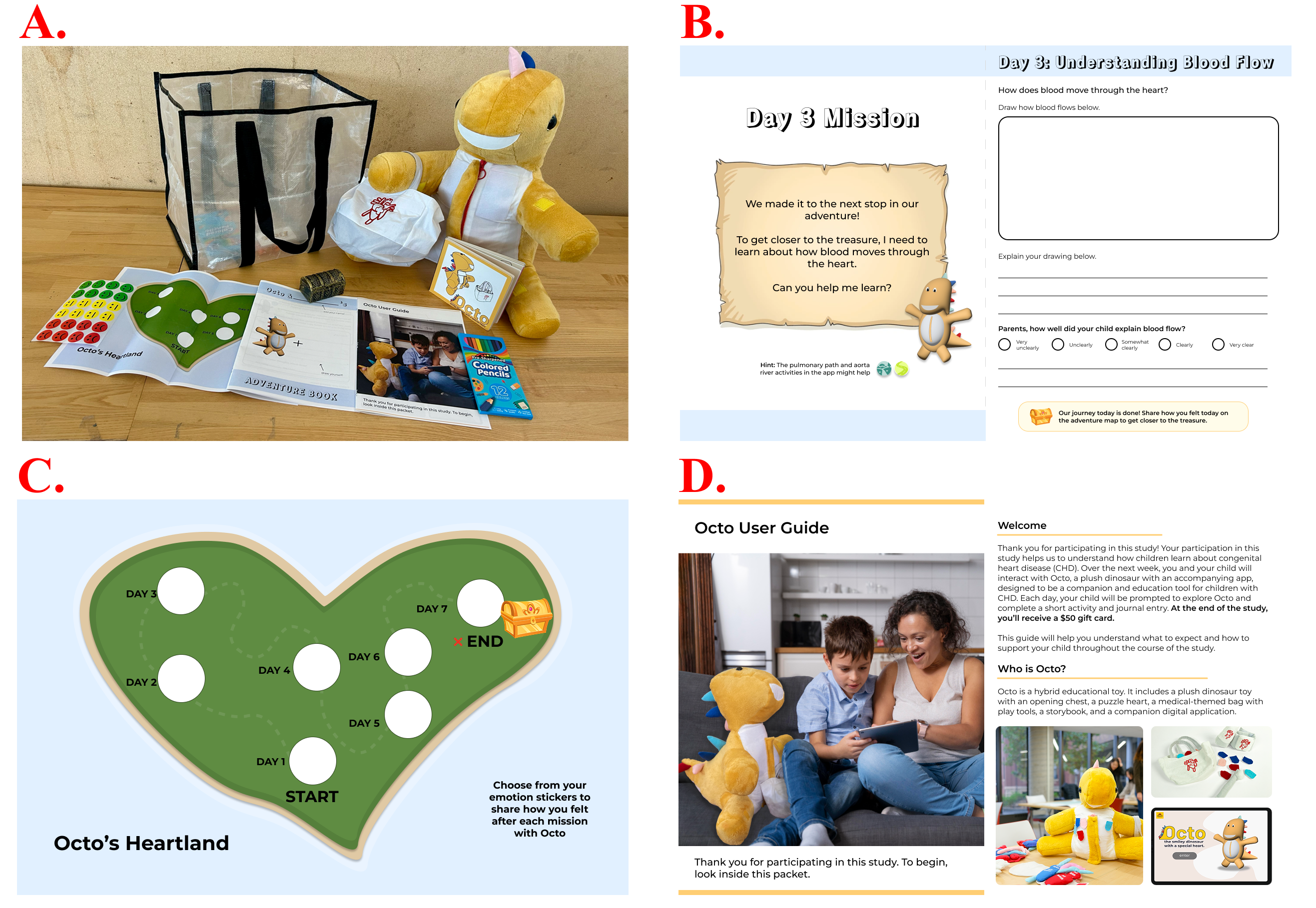}
  \caption{Deployment kit, including an activity journal, an emotional journey map, and a parent user guide.}
  \Description{Fig 1. Image shows materials for the Octo at-home deployment kit in four parts. Part A shows an image of the entire at-home deployment kit laid out. From left to right, there are Likert-scale emotion stickers, an emotion tracking map that has an area to log emotions for each day, an activity journal, a parent user guide, a box of colored pencils, the storybook which includes an illustration of Octo on the cover, and the physical prototype of Octo, yellow in color, carrying the medical bag which includes an illustration of the heart on its front. Part B shows a two-page spread of the activity journal; there is a blue header which includes the learning goal for the day and a mission statement on the left page with an illustration of Octo in a speech bubble, the right side has an open area for drawing and lines to write responses to questions. Part C shows the emotional logging map; this is a green heart-shaped island in a blue ocean. There are seven empty circles on this map meant to be filled by the Likert-scale emotion stickers. Each circle is labeled with a day number to keep track. Part D shows the cover and welcome page of the parent user guide; the cover page shows a parent and child sitting on a couch playing with Octo, while the welcome page shows text describing the background of the study and three pictures of the Octo prototype. The first picture shows the Octo prototype laid out, the second shows the medical bag and its tools, and the third shows a mock up of the digital app on a tablet screen.}
  \label{fig:Fig1}
\end{figure}

After obtaining parental consent, child assent, and completing pre-assessments, we distributed an initial incentive, provided the deployment kit to families, and scheduled a return appointment. The deployment kit included (a) the Octo physical prototype, (b) a 7-day activity journal, (c) an emotion-tracking map with emotion stickers, colored pencils, a treasure chest reward, and (d) a user guide with access to the digital prototype, all stored in a portable tote bag (Figure~\ref{fig:Fig1} and Supplemental Material). We designed the kit to support child-centered feedback during everyday use by accommodating children's developmental needs and incorporating insights from prior phases \cite{Barbazi2025Assessment, Barbazi2026}. Younger children participated with parental assistance limited to scribing rather than responding on their behalf, whereas older children participated independently. Children provided feedback through drawing, writing, or verbal expression, allowing them to communicate in developmentally appropriate ways. Parents could also optionally add their own reflections. The journal used fictional inquiry \cite{Dindler2007} to frame each activity as a role-playing ‘mission’, creating emotional distance that supports open feedback \cite{Barbazi2025Play,Barbazi2026,Hiniker2017}. Daily missions targeted understanding CHD, healthy habits, and clinical procedures. Prompts were voiced through Octo to guide exploration (e.g., \textit{“What parts of my heart did you find?"}) and combined scavenger hunt tasks (e.g.,\textit{"Circle the healthy habits you found"}), drawing prompts (e.g., \textit{"Draw how blood flows"}), and short reflections (e.g., \textit{"What's your favorite thing you've learned?"}). In addition to journal responses, children logged their feelings using Likert‑style emotion stickers on the emotion-tracking map. In pediatric studies, symbolic feedback supported engagement \cite{Chen2026,Hong2020,Barbazi2026,Read2006} and stickers allowed us to track emotional feedback. \enlargethispage*{16pt}

Completing daily activities moved children closer to unlocking the treasure chest reward, which supported pediatric engagement through simple gamification \cite{Teela2023, Zeng2025}, and completion of the mission \cite{Dindler2007}. The research team remained available for troubleshooting. At the end of the week, families returned the kit at the Explorer Clinic, completed post-assessments, and received a final incentive. For two families, we conducted the return process at their homes due to scheduling needs.  

\subsection{Data Collection and Analysis} 

We collected completed activity journals, daily emotion-tracking maps, parent user guides, and photographs that parents voluntarily shared. Many parents shared informal reflections during return sessions, which provided additional contextual understanding but were not coded. We digitized all materials in Miro (a collaborative, digital workspace) by scanning artifacts, transcribing written feedback into digital notecards, and recreating emotion-tracking maps. We then coded visual and written feedback, including children’s drawings and photographs, to create cross-referenced datasets for each family. After coding all participant data, we used reflexive thematic analysis to refine and group related codes into emerging themes across the dataset \cite{Clarke2017}. \enlargethispage*{16pt}

\section{Results}
\subsection{Children’s Learning Behaviors and Health Literacy}
\begin{figure}[h]
  \centering
  \includegraphics[width=\linewidth]{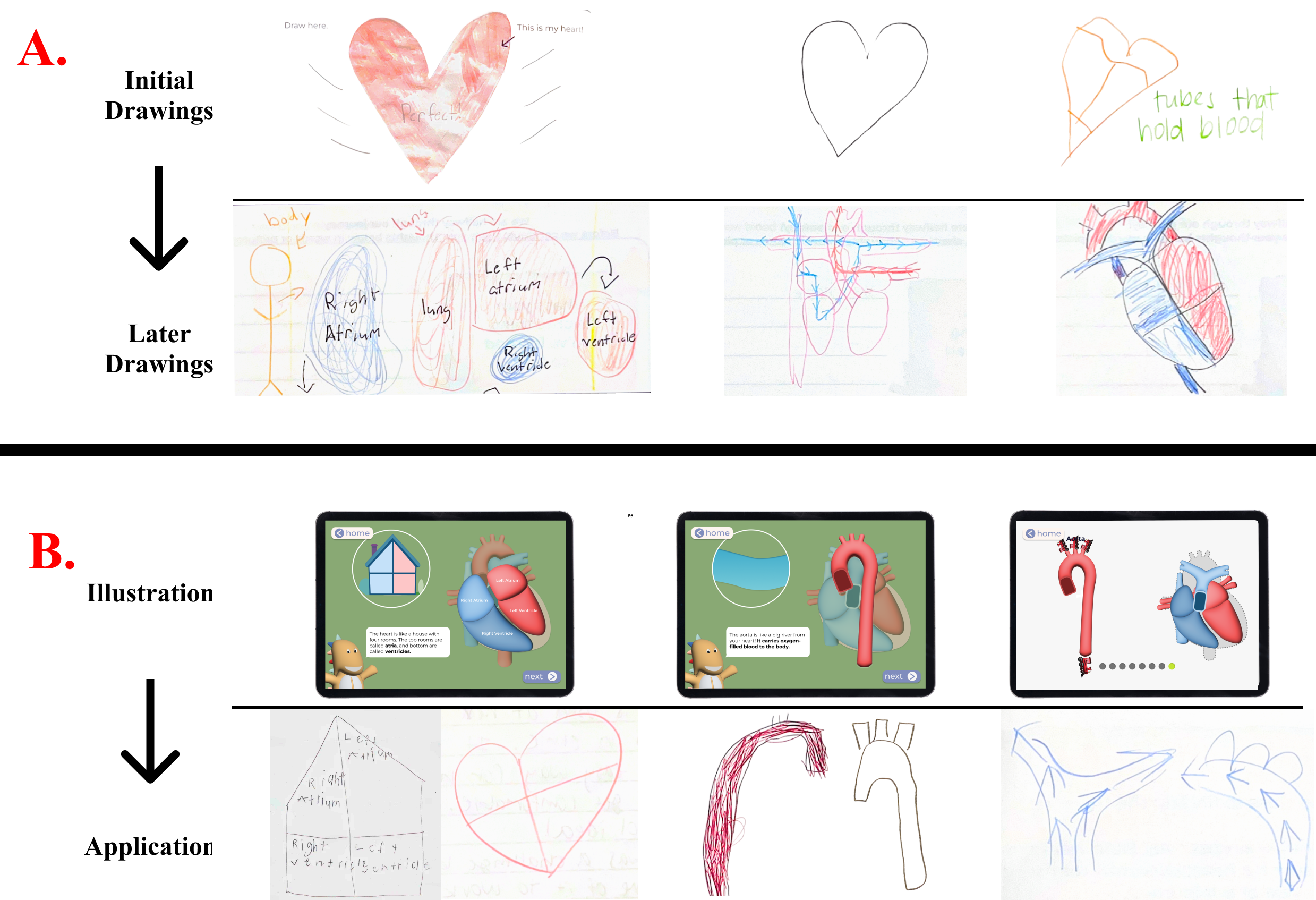}
  \caption{Children’s drawings of their hearts before/after using Octo and applications of illustrations}
  \Description{Fig 2. has two parts. Part A shows three drawings of simple hearts drawn by children. Under these drawings are drawings completed by the same child after using Octo. These drawings display images of anatomically complex hearts with labels for blood flow, vessels, and chambers colored in red and blue pencil. On the left of these rows is an arrow pointing down from the text 'Initial Drawings' to the text 'Later Drawings'. Part B shows three screens from the app in one row and drawings from children in a row below. The first screen shows an illustration of a house compared to the four chambers of the heart. Under this illustration are two examples of children drawing representations of the heart similar to this illustration. The second illustration shows an illustration of the aorta in the app. Under this are two drawings from children that depict the aorta accurately. The third illustration is a screen from the app showing how blood flows through the aorta. The children's drawing under this illustration shows arrows of blood flow through the aorta that match what is depicted in the image of the app above. On the left of these rows is an arrow pointing down from the text 'Illustration' to the text 'Application'. }
  \label{fig:Fig2}
\end{figure}

Children demonstrated a progressive understanding of heart function despite initial differences in baseline understanding due to age and health conditions. Early drawings depicted simple symbolic hearts, while later drawings showed complex anatomical features, such as blood flow, chambers, and vessels (Figure~\ref{fig:Fig2}a). Visual and narrative aids supported children’s understanding, as many children referenced illustrations from the prototype when drawing or explaining the heart (Figure~\ref{fig:Fig2}b). For example, P3 said the heart \textit{“has an upstairs and downstairs, and the blood travels from top to bottom”}, and P11 wrote, \textit{“blood flows through the body like a river”}. Parents noticed their children’s improved understanding. P8 was surprised at \textit{“how quickly [her child] learned to put the heart together and retained the information”}. While children gained conceptual understanding of the heart’s components and functions, specific terminology remained challenging for several children, requiring parental prompting (P4, P6, P7, P8). Children’s comprehension was supported by the hybrid play of Octo’s digital and physical prototypes. C10 changed his journal response about blood flow after checking the app for clarification, and C7 and C8 used the heart model, storybook, and app to match heart pieces to their names. After playing with the medical bag, most children could visually identify tools and describe their functions. Several parents suggested aligning the digital app with the clinical elements of the physical prototype, as this was a strong area of engagement.

\subsection{Dynamic Range of Children’s Emotional Expression}
\begin{figure}[h]
  \centering
  \includegraphics[width=\linewidth]{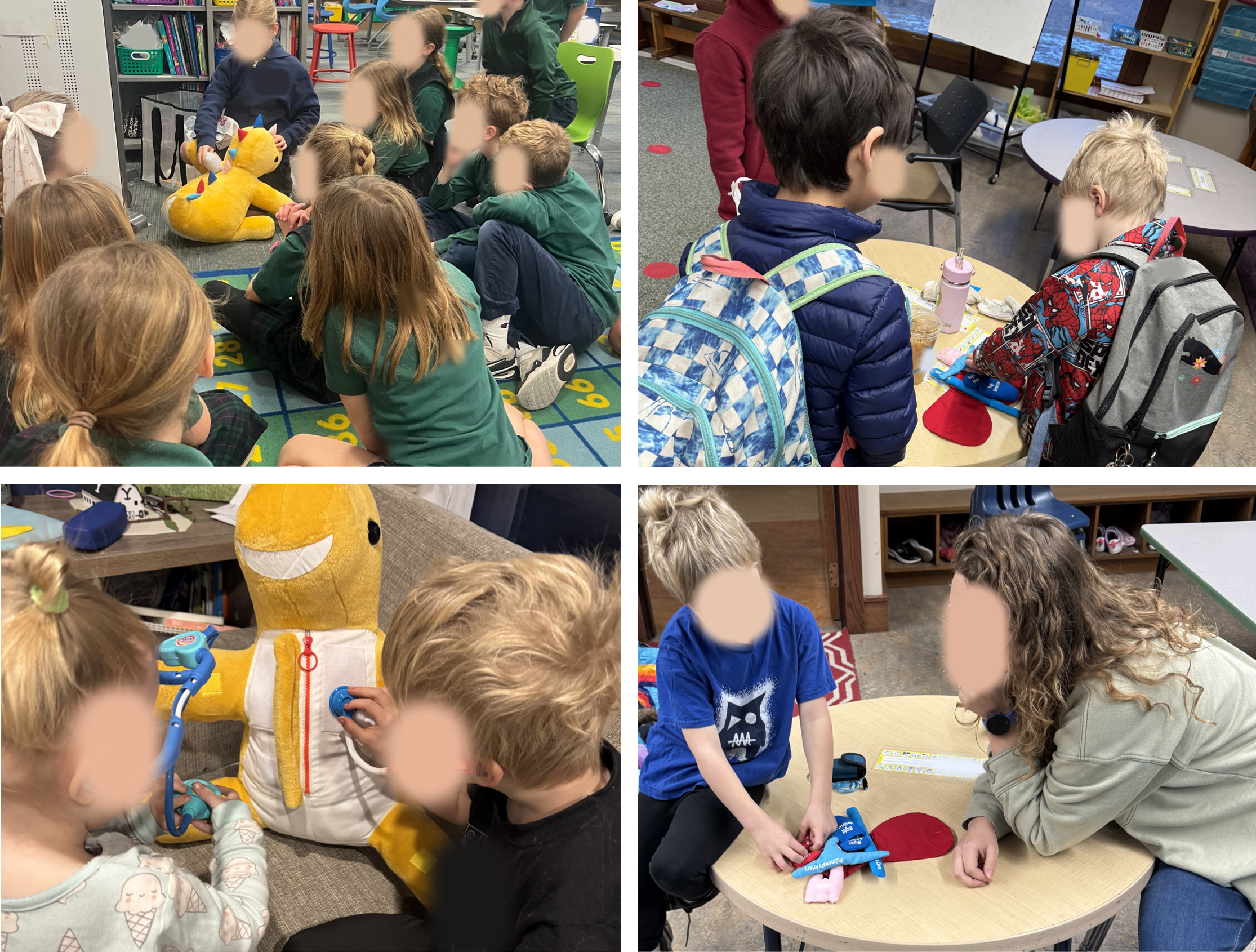}
  \caption{Children showing Octo to classmates, siblings, and teachers.}
  \Description{Fig 3. shows four pictures of children with Octo in different settings. From left to right, the first image shows a child showing the physical Octo prototype to their class during show and tell. The children in the class look at Octo with interest. The second image shows a child showing two classmates the puzzle heart from the physical Octo prototype. Both classmates look at the heart while the main child participant displays it on a table. The third picture shows a child with their sibling playing with Octo using toy stethoscopes. The fourth image shows a child showing the puzzle heart to their teacher.}
  \label{fig:Fig3}
  
\end{figure}

Children displayed confidence, as many expressed pride in their growing understanding, which motivated continued engagement. C6 \textit{“felt proud of remembering [the function of] the aorta” }(P6) and C9 shifted from feeling frustrated about \textit{“not knowing enough”} (P9) to wanting to learn more about the heart to \textit{“teach other kids”}. After memorizing all the parts of the heart, C5 explained the heart \textit{“to anyone that would listen”} (P5). Several children displayed confidence by bringing Octo to school and sharing what they learned with classmates, while others taught their siblings or extended family (Figure~\ref{fig:Fig3}). Many parents noted this was a surprise, as P6 said, \textit{"she talks about her heart like she’s the expert now... she's so confident talking about her heart"}. Learning fostered confidence in having a \textit{“special heart”} (C3, C6, C9).

Emotional attachment also influenced children’s engagement with the tool throughout the week. Some kids viewed Octo as their \textit{"best friend"} (C7, C9, C13), the \textit{“best toy ever”} (C10), or expressed love for the toy (C3, C5, C6, C8). This sense of companionship shaped emotional responses: when C13 couldn’t play with Octo for part of the day, she recorded a negative emotion on the emotion map, and many children expressed sadness when returning Octo (C9, C13, C5, C3). Emotional patterns in feedback showed engagement is more related to attachment than to age. C1 and C4, both older, initially struggled to focus. As the week progressed, C4’s feedback increased due to developing attachment, while C1 did not form attachment and recorded more negative emotions.\enlargethispage*{16pt}

\subsection{Impacts on Parent-Child Communication} 
Parents consistently described Octo as a facilitator for conversations about their child’s heart condition. While parents still led conversations, Octo guided conversations in an approachable manner through simple terms in the storybook and app (P5, P9), and the interactive visual reference of the physical heart (P3, P7, P5, P6, P8). P8 stated \textit{“using Octo, the storybook, and app allowed [them] to have a detailed talk”} about their child’s heart that had never occurred before, and P3 expressed that Octo created \textit{“a shared space for open conversation”}. P11 discussed a past surgery in depth with her child for the first time without fear of bringing up past trauma. Facilitating conversations that had previously been too difficult or overwhelming had an emotional impact. Many parents found it \textit{"heartwarming"} to witness their child’s excitement and growth and were moved to tears upon reflection (P3, P5, P7). At the same time, usability issues with the app occasionally disrupted the learning process and required parental intervention. \enlargethispage*{16pt}
 
\section{Discussion, Future Work, and Conclusion}
Deploying pediatric health technology in home settings generated insights into use that differed from those produced by prior studies focused on clinical evaluations, collaborative tracking systems, or pre-post assessments \cite{Barbazi2025Scoping,Cha2024,Cha2025}. Prior pediatric interventions have examined family communication, collaborative care management, and caregiver understanding through structured assessments or caregiver reports \cite{Cha2024,Rodts2020,Shin2019}. Our deployment captured how these interactions unfolded within everyday routines and family life, showing that children’s understanding developed gradually as familial involvement encouraged repeated engagement with Octo’s physical and digital components. This is consistent with HCI literature that frames children’s health literacy and health management as ongoing, family-embedded processes rather than isolated events \cite{Richards2023,Su2023}. Prior research emphasizes collaborative reflection, routines, and children’s growing involvement in everyday health practices \cite{Nikkhah2022Care,Nikkhah2022Adaptive}; our findings show how educational interventions can support iterative learning and communication. 

Furthermore, our deployment revealed how educational tools can mediate collaborative family communication beyond clinical encounters. While prior technologies have supported caregiver coordination, information management, and condition tracking \cite{Cha2024,Sepehri2023}, Octo facilitated previously difficult conversations in daily life. Parents described Octo as making complex medical topics more approachable and enabling shared learning. This contributes to CSCW research on distributed caregiving and family communication \cite{Nikkhah2022Care,SantosSilva2025,Shin2018} by demonstrating how pediatric educational interventions can facilitate communication about complex conditions in everyday life. Finally, we highlight the value of extending PD into deployment. We demonstrated a new application of established methods to pediatric deployments by applying participatory approaches commonly used in early design stages (e.g., fictional inquiry, drawing, diary probes) \cite{Druin1999,Isola2012,Dindler2007} to the mixed-methods practices of healthcare deployments. While multimodal feedback collection is prevalent in pediatric deployments, through surveys, interviews, and check-ins, direct feedback from children remains limited \cite{Moon2025,Ullman2021, Park2022}. Our participatory feedback approach helped capture children’s direct experiences rather than relying solely on parental mediation or structured post-study evaluations. This study suggests how future deployments can better capture children’s engagement during everyday use beyond standard pre-post assessments.

Although limited by a short timeframe and a relatively homogeneous participant group, this study advances HCI and health research on pediatric technologies by showing how field deployment can reveal everyday learning, communication, and caregiving practices that structured evaluations may overlook. Future phases of this research will refine Octo's physical and digital components and extend deployment into clinical settings to further examine how communication, learning, and caregiving practices intersect to shape children’s experiences across home and clinical settings. \enlargethispage*{12pt}

\begin{acks}
We would like to thank participants, providers, and staff at M Health Fairview Explorer Clinic and Camp Odayin. Appreciation to former students: Hannah Gootzeit, Irene Zeng, Grace Rubas, Jessica Espinosa, Jonathan Jakubas, Levi Skelton, Andy Thai. This research was supported by the Lasting Imprint Foundation and the University of Minnesota’s Research Opportunities Program (UROP).
\end{acks}

\bibliographystyle{ACM-Reference-Format}
\bibliography{sample-base}

\end{document}